\documentclass[aps,prl,twocolumn,floats,showpacs,floatfix,a4paper]{revtex4}
\usepackage{epsfig}
\usepackage{color}
\usepackage{bm}
\usepackage{latexsym}
\begin{document}
\newcommand{\cc}{{\bf\Large C }}
\newcommand{\hide}[1]{}
\newcommand{\tbox}[1]{\mbox{\tiny #1}}
\newcommand{\half}{\mbox{\small $\frac{1}{2}$}}
\newcommand{\sinc}{\mbox{sinc}}
\newcommand{\const}{\mbox{const}}
\newcommand{\trc}{\mbox{trace}}
\newcommand{\intt}{\int\!\!\!\!\int }
\newcommand{\ointt}{\int\!\!\!\!\int\!\!\!\!\!\circ\ }
\newcommand{\eexp}{\mbox{e}^}
\newcommand{\EPS} {\mbox{\LARGE $\epsilon$}}
\newcommand{\ar}{\mathsf r}
\newcommand{\im}{\mbox{Im}}
\newcommand{\re}{\mbox{Re}}
\newcommand{\bmsf}[1]{\bm{\mathsf{#1}}}
\newcommand{\dd}[1]{\:\mbox{d}#1}
\newcommand{\abs}[1]{\left|#1\right|}
\newcommand{\bra}[1]{\left\langle #1\right|}
\newcommand{\ket}[1]{\left|#1\right\rangle }
\newcommand{\mbf}[1]{{\mathbf #1}}
\newcommand{\eos}{\,.}
\definecolor{red}{rgb}{1,0.0,0.0}

\title{ {\bfseries Critical Fidelity at the Metal-Insulator Transition} }

\author{
Gim Seng Ng$^{1,2}$, Joshua Bodyfelt$^{1}$ and Tsampikos Kottos$^{1,2}$
}

\affiliation{
$^1$Department of Physics, Wesleyan University, Middletown, Connecticut 06459, USA \\
$^2$Max-Planck-Institute for Dynamics and Self-Organization, Bunsenstra\ss e 10, D-37073 G\"ottingen, Germany
}

\begin{abstract}
Using a Wigner Lorentzian Random Matrix ensemble, we study the fidelity, $F(t)$, of systems 
at the Anderson metal-insulator transition, subject to small perturbations that preserve the 
criticality. We find that there are three decay regimes as perturbation strength increases: 
the first two are associated with a gaussian and an exponential decay respectively and can be 
described using Linear Response Theory. For stronger perturbations $F(t)$ decays algebraically 
as $F(t)\sim t^{-D_2^{\mu}}$, where $D_2^{\mu}$ is the correlation dimension of the Local Density of States.
\end{abstract}
\maketitle


The theory of fidelity \cite{P84} (also known as Loschmidt Echo) has been a subject of intensive 
research activity during the last years (for a recent review see \cite{GPSZ06}). This interest 
has been motivated by various areas of physics, ranging from atomic optics \cite{GCZ97,AKD03,
KASDMRKGRM03}, microwaves \cite{SGSS05} and elastic waves \cite{LW03} to quantum information 
\cite{NC00} and quantum chaos \cite{JP01,JSB01,JAB04,CT02,PS02,BC02,WC02,PLU95,VH03,CLMPV02}. 
It has been adopted as a standard measure for quantum reversibility and stability of quantum 
motion with respect to changes in an external parameter $x_{\rm e}$. Formally, the fidelity $F(t)$, 
is defined as:
\begin{equation}
\label{eq:FidDef}
F(t) = |\bra{\psi_0}\eexp{iH_{\rm b}t}\eexp{-iH_{\rm f}t}\ket{\psi_0}|^2 ;\quad \hbar=1
\end{equation}
where ${\bf H}_{\rm f}$ and ${\bf H}_{\rm b}= {\bf H}_{\rm f}+x_{\rm e}{\bf B}$ represent the 
reference Hamiltonian and its perturbed variant, respectively, while $\ket{\psi_0}$ is 
an initial state. One can interpret fidelity (\ref{eq:FidDef}) in two equivalent ways. 
It can be considered as the overlap of an initial state with the state obtained after the 
forward unperturbed evolution, followed by a backward perturbed evolution. Equivalently, 
it is the overlap of a state obtained after a forward unperturbed evolution and the state 
after a forward perturbed evolution. The latter interpretation is closely linked to the 
concept of dephasing \cite{Z91} in mesoscopic devices and coherent manipulation of a 
quantum state. Sustaining the coherence of a superposition of state vectors is at the 
heart of quantum parallelism in quantum computation schemes \cite{SAI89,FH03,NC00}. 
The first interpretation goes back to the original proposal by Peres \cite{P84}, who 
used fidelity to study quantum-classical correspondence and identify traces of classical 
(chaotic or integrable) dynamics in quantized systems. 

For a quantum system with a classical chaotic counterpart, the decay of the fidelity depends 
on the strength of the perturbation parameter $x_{\rm e}$. Recent studies indicated that 
there are three $x_{\rm e}-$regimes: the standard perturbative regime, the Fermi Golden Rule 
regime (FGR), and the non-perturbative regime. The first two can be described by Linear Response 
Theory (LRT) 
leading to a decay which depends on the perturbation strength $x_{\rm e}$ as $F(t)\sim\eexp{
-(x_{\rm e}t)^2}$ and $F(t)\sim \eexp{-x_{\rm e}^2t}$, respectively \cite{PS02,JSB01}. 
In the non-perturbative regime, the decay is $F(t)\sim \eexp{-\lambda t}$, with a rate that 
is perturbation independent and is given by the Lyapunov exponent $\lambda$ of the underlying 
classical system \cite{JP01,CLMPV02,JSB01}.

The investigation of the fidelity has recently been extended to systems that have 
integrable classical dynamics. It was shown \cite{JAB04} that the decay follows a power 
law $F(t)\sim t^{-3d/2}$, where $d$ is the dimensionality of the system. A similar algebraic 
decay was found for disordered systems with diffractive scatterers, where now the power 
law is governed by the diffusive dynamics \cite{AGM03}.

Despite the progress in understanding the fidelity of various 
systems, a significant class was left out of the investigation. These are systems which
show an Anderson metal-insulator transition (MIT) as an external parameter changes. In 
the metallic regime, the eigenstates of these systems are extended, and the statistical 
properties of their spectrum are quite well described by random matrix theory~\cite{SSSLS93}. 
In particular, the level spacing distribution is very well fitted by the Wigner surmise. 
Deep in the localized regime, the levels become uncorrelated leading to a Poissonian level 
spacing distribution and the eigenfunctions are exponentially localized. At the MIT, the 
eigenfunctions are critical, exhibiting multifractal structure characterized by strong 
fluctuations on all scales. The eigenvalue statistics are characterized by a third 
universal distribution \cite{SSSLS93,AS86}. Representatives of this class are disordered 
systems in $d>2$ dimensions, two-dimensional systems in strong magnetic fields (quantum 
Hall transition), or periodically kicked systems with a logarithmic potential singularity 
\cite{GW05}. 

Here, for the first time, we address the behavior of $F(t)$ for systems at criticality and 
present consequences of the MIT on the fidelity decay. Using the Wigner Lorentzian Random
Matrix (WLRM) ensemble, we find that there are three regimes: (a) the standard perturbative 
regime where the decay is gaussian; (b) the FGR decay where the decay is exponential and 
(c) the non-perturbative regime where an initial gaussian decay (Zeno decay) is followed 
by a power law. The latter decay is novel and reflects the critical nature of the system. 
Specifically we found that
\begin{equation}
\label{Fcrit}
F(t)\sim {1\over t^{D_2^{\mu}}}
\end{equation}
where $D_2^{\mu}=D_2^{\psi}/d$ \cite{HK97} is the correlation dimension of the Local Density 
of States (LDoS) while $D_2^{\psi}$ is the correlation dimension of the critical eigenstates 
and $d$ is the actual dimensionality of the system. For the WLRM model $D_2^{\mu} = D_2^{\psi}
=D_2$ since $d=1$. The correlation dimension $D_2^{\psi}$ is usually defined through the 
inverse participation ratio, $P_2=\int d^dr|\psi(r)|^{4}\sim L^{-D_2^{\psi}}$, where $L$ is 
the size of the system \cite{W80}. The correlation dimension is also related to the spectral 
compressibility $\chi=(d-D_2^{\psi})/2d$, defined through the level number variance $(\delta 
N)^2 \approx \chi \langle N\rangle$ \cite{CKL96,EM00,ME00}. At the same time, $D_2^{\psi}$ 
manifests itself in a variety of other physical observables. As examples, we mention the 
conductance distribution \cite{BHMM01,P98}, the anomalous spreading of a wave-packet 
\cite{KKKG97}, the spatial dispersion of the diffusion coefficient \cite{CD88,C90,HK99}, 
and the anomalous scaling of Wigner delay times \cite{MK05}.


We use the WLRM model \cite{MFDQS96,KT00,MKC06a}, defined as:
\begin{equation}
\label{eq:WLRMDef}
{\bf H}= {\bf H_0} + x{\bf B}
\end{equation}
Both  $\bf{H_0\mbox{ and }B}$ are real symmetric matrices of size $L \times L$ with matrix
elements randomly drawn from a normal distribution with zero mean and a variance depending 
on the distance of the matrix element from the diagonal
\begin{equation}
\label{eq:BVar}
\langle \sigma_{nm}^2\rangle = \frac{1}{1+|\frac{n-m}{b}|^2}.
\end{equation}
where $b\in (0,L)$ is a free parameter that controls the critical properties of the system
(see Eq.~(\ref{Dq}) below). Random matrix models with variance given by (\ref{eq:BVar}) were 
introduced in \cite{MFDQS96} and further studied in \cite{M00,KT00,EM00,V03}. Field-theoretical 
considerations \cite{MFDQS96,M00,KT00} and detail numerical investigations \cite{EM00,V03} 
verify that the model shows all the key features of the Anderson MIT, including multifractality 
of eigenfunctions and non-trivial spectral statistics at the critical point. A theoretical 
estimation for the correlation dimension $D_2^{\psi}$ gives \cite{MFDQS96}
\begin{equation}
\label{Dq}
D_2^{\psi} =\left\{
\begin{array}{cc}
 4b \Gamma(3/2)[\sqrt{\pi}\Gamma(1)]^{-1} \ &, b \ll 1 \nonumber\\
 1-2(2\pi b)^{-1} \   &, b\gg 1
\end{array}
\right.
\end{equation}
where $\Gamma$ is the Gamma function. 

\begin{figure}
\includegraphics[width=\columnwidth,keepaspectratio,clip]{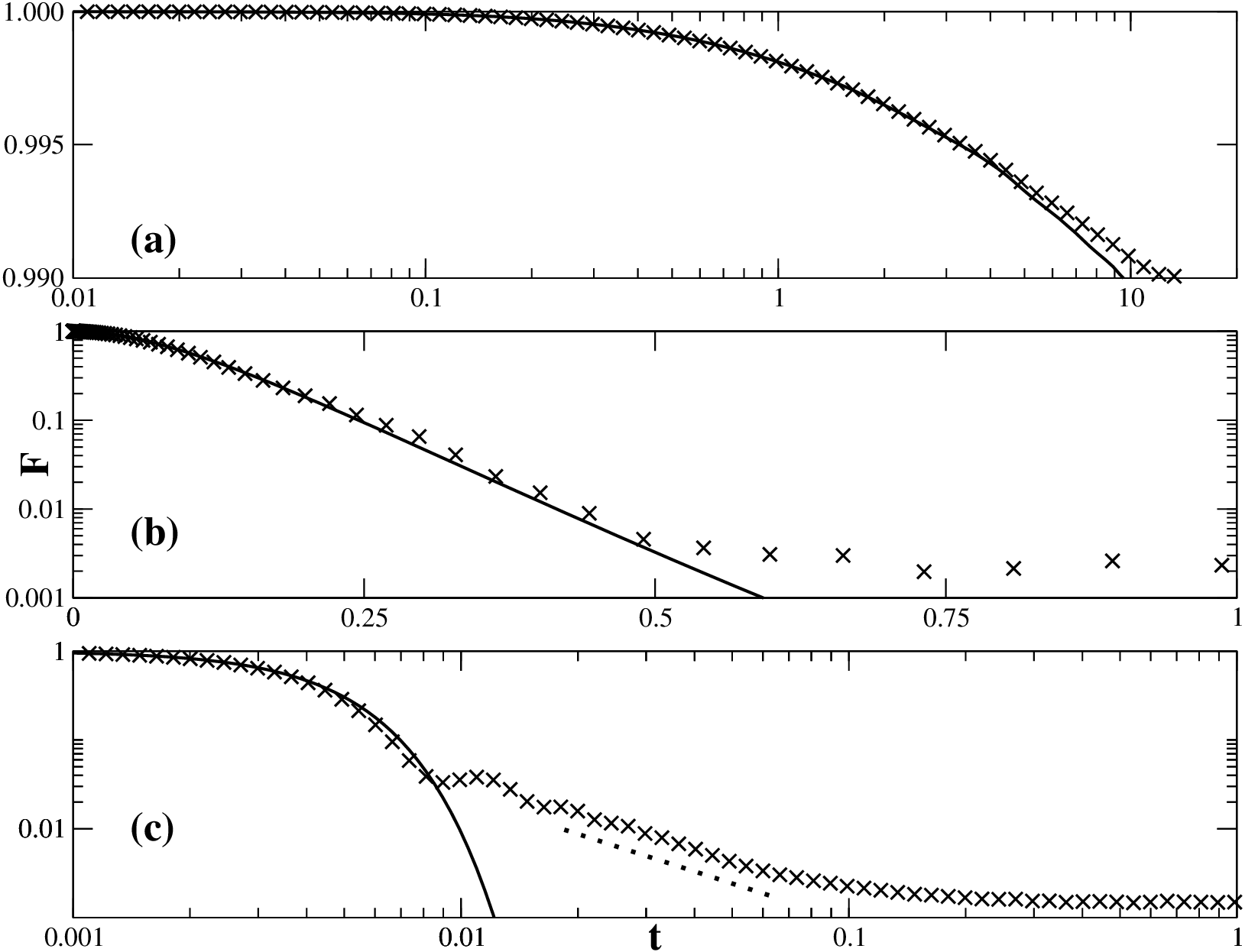}
\caption{\label{cap:fig1}
Fidelity of an ES, for (a) $x=0.01$ (the standard perturbative regime), (b) $x=0.8$ (FGR regime) 
and (c) $x=20$ (non-perturbative regime). The solid lines are the LRT results from Eq.~(\ref{eq:FidExp}) 
while the crosses are the outcomes of the numerical simulations with model (\ref{eq:WLRMDef},
\ref{eq:BVar}). In these simulations $L=1000$ and $b=10$. The mean level spacing of the 
unperturbed system is set to $\Delta\approx 1$. In 
this case, $x_c \approx 0.59$ and $x_{prt} \approx 1.88$. The dotted line in (c) is 
plotted to guide the eye on the power-law behavior.
}\end{figure}
The forward and backward Hamiltonians used for the calculation of the fidelity
(\ref{eq:FidDef}) are \cite{note1}
\begin{equation}
\label{fb}
{\bf H_{\rm f}} = {\bf H}(x)\quad\quad {\rm and}\quad\quad {\bf H_{\rm b}}= {\bf H}(-x)
\end{equation}
We operate in the basis where ${\bf H_0}$ is diagonal \cite{MKC06a}. In this basis, the 
perturbation matrix ${\bf B}$ is $x-$invariant \cite{MKC06a}, i.e. it preserves the same 
Lorentzian power-law shape (\ref{eq:BVar}), while its critical properties (like the multifractal 
dimension $D_2^{\psi}$) remain unchanged. For the numerical evaluation of $F(t)$, we have 
used two types of initial conditions $\ket{\psi_0}$: an eigenstate of ${\bf H_0}$ (ES) and 
a generic ``random" state (RS). In both cases, the results are qualitatively the same. 
Therefore, we will not distinguish between them. In our numerical experiments we used 
matrices of size varying from $L=1000$ to $L=5000$. We have performed an averaging over 
different initial states and realizations of the perturbation matrix ${\bf B}$ (typically 
more than $1000$). 


An overview of the temporal behavior of the fidelity $F(t)$ for three representative 
perturbation strengths is shown in Fig. 1. For perturbation strengths smaller than $x_c 
\approx \frac{\Delta}{\sqrt{\pi}}\sqrt{1+\frac{1}{b}}$ \cite{note2}, the decay of $F(t)$ is 
gaussian (see Fig. 1a). The perturbative border $x_c$ is the perturbation strength needed 
in order to mix levels within a distance of a mean level spacing $\Delta$ \cite{MKC06a}. 
Above this border, one typically expects an exponential FGR decay of fidelity \cite{JSB01}, 
with a rate given by the width of the Local Density of States (LDoS) \cite{MKC06a} (see 
Fig. 1b). We can apply LRT \cite{GPSZ06} to evaluate the decay of $F(t)$ in these two 
regimes. The resulting expression reads
\begin{equation}
\langle F(t)\rangle_{B,n_0} \approx 1-(2x)^2 {\cal C}(t) \approx
\eexp{-(2x)^2{\cal C}(t)} \label{eq:FidExp}
\end{equation}
where $\langle\ldots\rangle_{B,n_0}$ represents a double average over $\bf{B}$ and initial 
states. The right hand side of expression (\ref{eq:FidExp}) assumes the
validity of infinite order perturbation theory. The correlator ${\cal C}(t)$ is 
\begin{equation}
{\cal C}(t)=
\int_0^t d\tau_1 \int_0^{\tau_1} d\tau_2 \sum_n |c_n|^2 \tilde{C}_n (\tau_1 - \tau_2)
- 2 {\cal I} t^2 \label{eq:FidExp1}
\end{equation}
where ${\cal I}=\sum_n |c_n|^4$ is the inverse participation ratio of the initial state,
$\tilde{C}_n (t - t')\equiv 2 (1+ \sum_{\gamma} \sigma^2_{n,\gamma} \cos[(E_{\gamma}^{(0)}-
E_n^{(0)})(t-t')])$, and $E^{(0)}_n$ denotes an eigenvalue of $\bf{H_0}$. In the case of 
standard GOE ensembles with $\sigma_{nm}^2 = 1$, Eq.~(\ref{eq:FidExp}) 
reduces to the expression derived in \cite{PS02}. The prediction of LRT (\ref{eq:FidExp}) 
is plotted together with the numerical results in Fig.~\ref{cap:fig1} for different 
perturbation strengths. A good agreement between Eq. (\ref{eq:FidExp}) and the numerical 
data is observed for perturbation strengths 
less than $x_{\rm prt}\approx \Delta \sqrt{b} \frac{\sqrt{ \pi- 1.28 [\pi/2-\mbox{arctan}
(1/b)]}}{1.68[\pi/2-\mbox{arctan} (1/b)]}$ (see Figs.~1a,b) \cite{note2}.


For $x$ larger than $x_{\rm prt}$  the decay of $F(t)$ cannot be captured by LRT (see Fig. 1c). 
The non-perturbative character of this regime was identified already in the frame of the parametric 
evolution of the Local Density of States (LDoS) \cite{MKC06a}. 
A representative temporal behavior of $F(t)$ for $x>x_{\rm prt}$ is reported in Fig. 1c.
For short times the decay of $F(t)$ is gaussian. For longer times, we can observe a
transition to a power law decay. The initial gaussian decay $F(t)\sim \eexp{-x^2t^2}$ is 
universal and can be identified with the quantum Zeno effect \cite{P84,GPSZ06}. It is valid
until times $t_Z\sim 1/x$. We will focus in the 
observed power-law decay which take place for $t>t_Z$. 

\begin{figure}
\includegraphics[width=\columnwidth,keepaspectratio,clip]{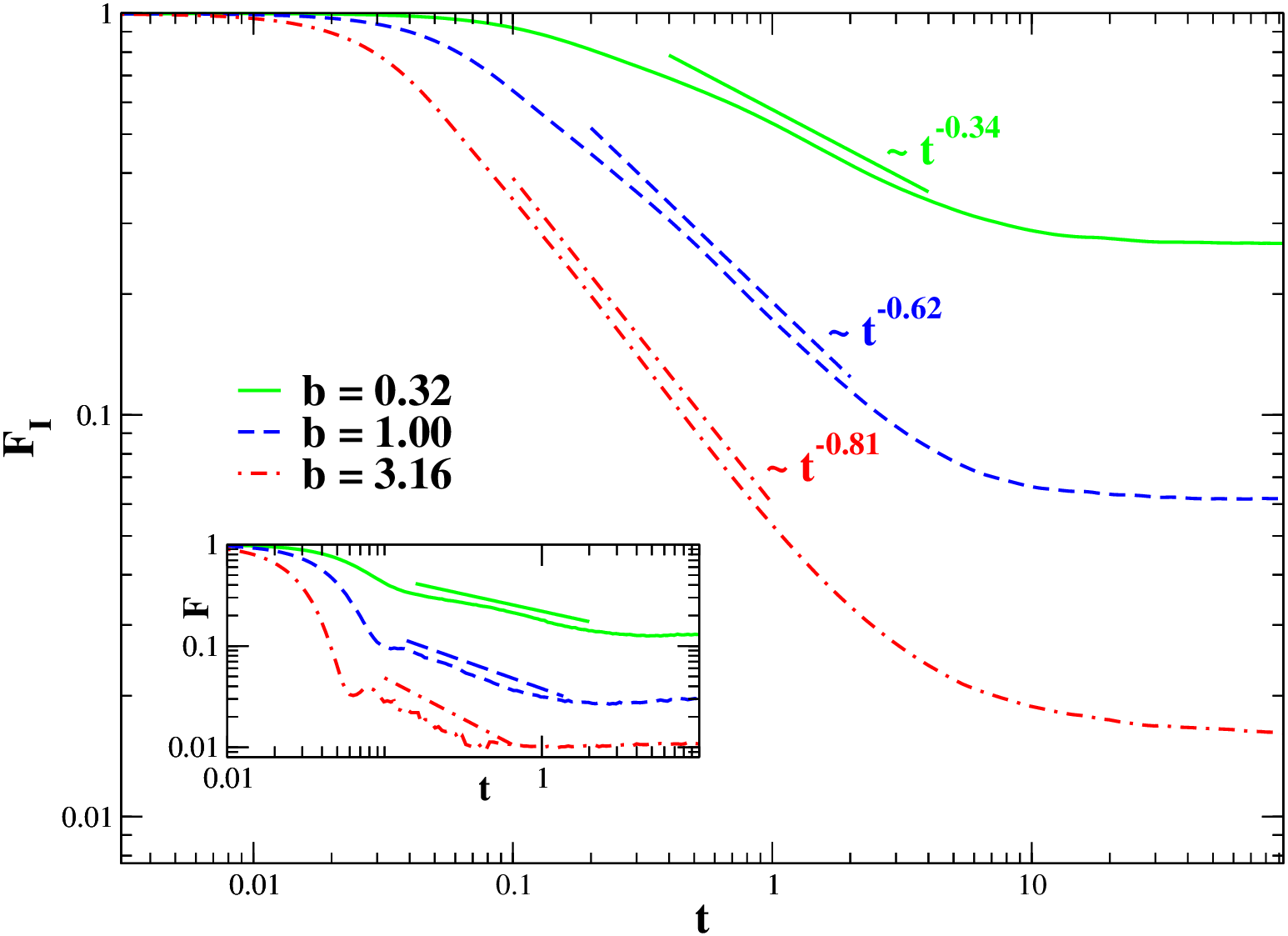}
\caption{\label{cap:fig2}(color)
$F_I$ for $b=0.32$, $1.00$ and $3.16$. The initial state was chosen to be an ES. The mean
level spacing of the unperturbed Hamiltonian is set to be $\Delta\approx 1$, while the 
size of the matrices is $L=5000$. In the cases reported here we have choose $x=5$. In the 
inset, we also present the fidelity for RS for $b=0.32$ and $b=1.00$ using the same parameters 
except $L=1000$. The straight lines are plotted to guide the eye.
}\end{figure}

\begin{figure}
\includegraphics[width=\columnwidth,keepaspectratio,clip]{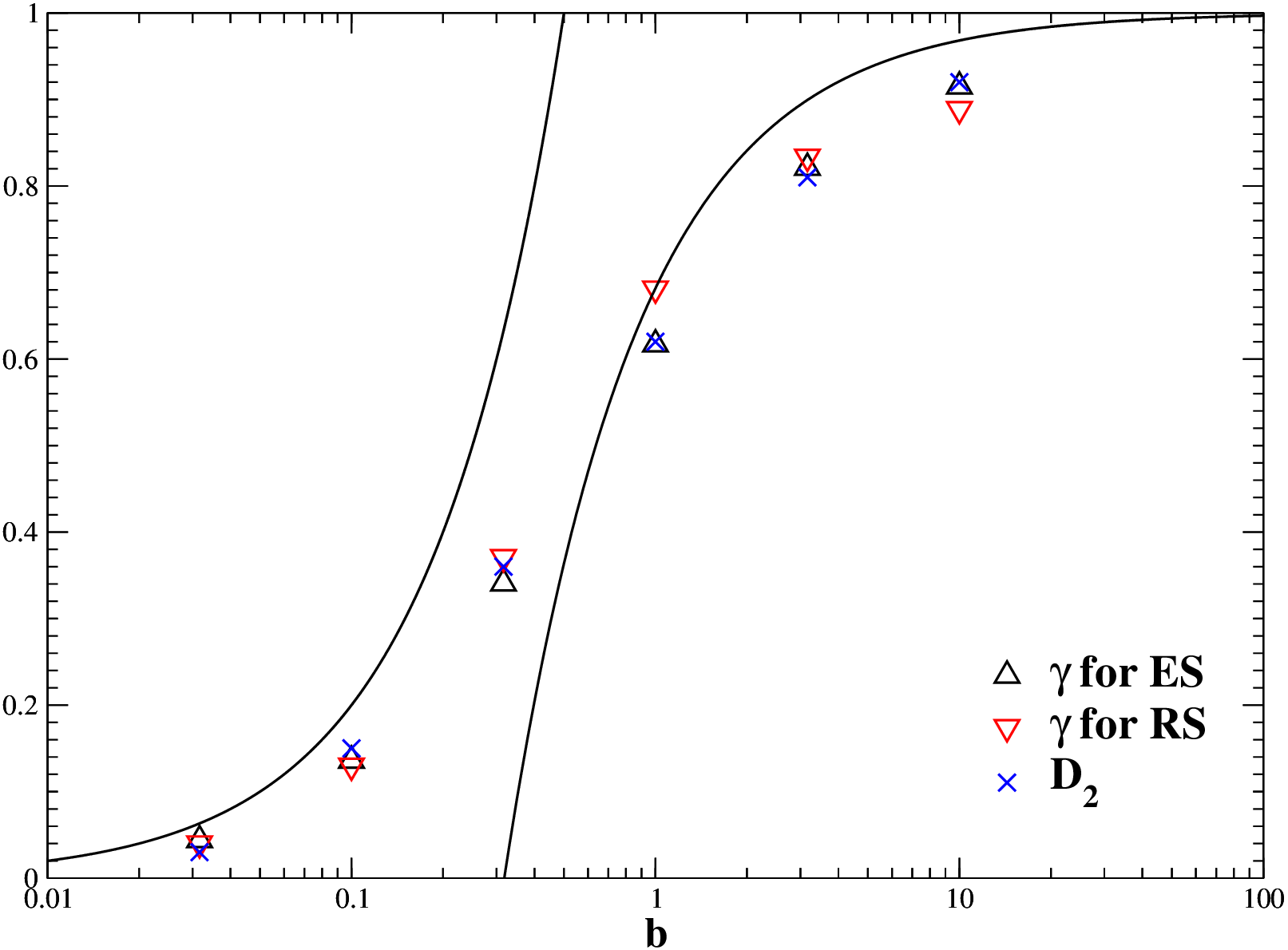}
\caption{\label{cap:fig3}(color)
The fitting parameter $\gamma$ for $b=0.03, 0.10, 0.32, 1.00, 3.16,$ and $10$. We are
using $\Delta=1$, $x=5$. The analytical (solid lines) Eq. (\ref{Dq}) and numerical (crosses)
\cite{M06} results for $D_2$ are also shown for comparison.
}\end{figure}

The numerical results for three different b-values, $b=0.32$, $1$, and $3.16$, are reported 
in Fig.~\ref{cap:fig2}. We use the time-averaged fidelity 
\begin{equation}
\label{FI}
F_I(t)\equiv\langle F(t)\rangle_t = \frac{1}{t}\int^t_0 F(t)dt
\end{equation}
to reduce further statistical fluctuations. In the inset, we present the raw data for 
the fidelity decay. In all cases the fidelity $F(t)$ clearly displays an inverse power law, 
\begin{equation}\label{eq:FI}
F(t) \propto \frac{1}{t^{\gamma}}
\end{equation}
with a power $\gamma$ that depends on the band-width parameter $b$. By fitting our data to 
Eq.~(\ref{eq:FI}), the power-law exponent $\gamma$ is extracted. In Fig.~\ref{cap:fig3} we 
summarize the extracted $\gamma$'s for both ES and RS initial conditions as a function of 
the bandwidth $b$. The results are essentially identical within the numerical accuracy of 
our fitting procedure. 

If the initial state $\ket{\psi_0}$ is an eigenstate of the backward (or forward) Hamiltonian 
then the fidelity is simply the survival probability $P(t)\equiv|\bra{\psi_0} \eexp{-iH_{\rm f}
t}\ket{\psi_0}|^2$ of wave-packet dynamics. In the latter case, it is known that the survival 
probability at criticality decays as $P(t)\sim 1/t^{D_2^{\mu}}$ \cite{KKKG97}. However, in these 
fidelity experiments, the initial state is neither eigenstate of $H_{\rm b}$ nor of $H_{\rm f}$. 
In fact, Ref.~ \cite{WC02} shows that the physics of quantum fidelity involves subtle cross 
correlations which in general are not captured by the survival probability (or the LDoS which 
is its Fourier transform) alone. Motivated by this equivalence between fidelity and survival 
probability for the specific choice of initial condition $\ket{\psi_0}$, we have compared in
Fig.~\ref{cap:fig3} the extracted power law exponents $\gamma$ with the correlation dimension 
$D_2^{\mu}=D_2^{\psi}=D_2$ \cite{V03}. The agreement between the $\gamma$ and the $D_2^{\mu}$ 
is excellent for all $b'$s confirming the prediction (\ref{Fcrit}). 

The connection between the exponent $\gamma$ and the fractal dimension $D_2^{\mu}$ calls for 
an argument for its explanation. The following heuristic argument provides some understanding 
of the power law decay Eq.~(\ref{Fcrit}). For any finite Hilbert space the fidelity $F(t)$ 
approaches the value $F_{\infty}=1/L$, being the inverse of the dimension of the Hilbert space. 
If the dynamics, however, take place in a space with an effective reduced dimension $D_2^{\psi}$, 
we will have $F_{\infty}=1/L^{D_2^{\psi}}$ \cite{note3}. Assuming a power law decay (\ref{eq:FI}) 
for the fidelity, we can estimate how the time $t_*$ at which $F(t_*)=F_{\infty}$ scales with
$L$, i.e. $t_*\sim L^{D_2^{\psi}/\gamma}$. On the other hand, the dynamics of a critical system 
is characterized by an anomalous diffusive law $L^2\sim t_*^{2D_2^{\mu}/ D_2^{\psi}}$ 
\cite{KKKG97} which defines the time $t_*\sim L^{D_2^{\psi}/D_2^{\mu}}$ needed to explore 
the available space $L$. Equating the two expressions for $t_*$ we finally get that 
$\gamma=D_2^{\mu}$. Although the numerical results leaves no doubt on the validity of Eq.~
(\ref{Fcrit}), a rigorous mathematical proof is more than desirable.

In conclusion, we have investigated the fidelity decay for systems at MIT. Depending on the 
perturbation strength $x$, we have indentified three distinct regimes: For $x<x_c$ the fidelity 
decay is gaussian; for $x_c<x<x_{\rm prt}$ the decay is exponential and for $x>x_{\rm prt}$ 
the decay is power law. The first two regimes are described by LRT. The third is non-perturbative. 
The power law decay is dictated by the critical nature of the system. Specifically, we have 
found that the power-law exponent is equal to the correlation dimension of the critical 
eigenstates.

We acknowledge T. Geisel for his continuous interest and support of this project. Useful
discussions with D. Cohen, T. Gorin, M. Hiller, S. Coppage, and A. Mendez-Bermudez are
also acknowledged.


\begin{thebibliography}{199}
\bibitem{note1} From Eq.~(\ref{eq:FidDef}) and Eq.~(\ref{fb}) we see that $x_{\rm e}=2x$.
\bibitem{note2} The values of $x_c$ and $x_{\rm prt}$ were derived in \cite{MKC06a} on the basis
of the LDoS analysis.
\bibitem{note3} For $x>x_{\rm prt}$ the wavefunctions are fractal and therefore fill only a 
fraction of the available space with an effective dimensionality given by $D_2^{\psi}$ 
\cite{MKC06a}.
\end{thebibliography}
\end{document}